\documentclass[prl,aps,twocolumn,amssymb,epsf,floatfix,floats,showpacs]{revtex4}
\usepackage{epsfig}
\newcommand{\be}{\begin{equation}}
\newcommand{\ee}{\end{equation}}
\newcommand{\bea}{\begin{eqnarray}}
\newcommand{\eea}{\end{eqnarray}}

\def\r#1{(\ref{#1})}

\def\i{{\rm i}}
\def\d{{\rm d}}
\def\e{{\rm e}}
\begin{document}

\title{Bethe Ansatz Solution of the Asymmetric Exclusion Process with Open
  Boundaries}
\author{Jan de Gier$^1$ and Fabian H.L. Essler $^2$}
\affiliation{$^1$ ARC Centre of Excellence for Mathematics and Statistics of
  Complex Systems, Department of Mathematics and Statistics, The University of
  Melbourne, 3010 VIC, Australia\\
$^2$ Rudolf Peierls Centre for Theoretical Physics, University
  of Oxford, 1 Keble Road, Oxford, OX1 3NP, United Kingdom}

\begin{abstract}
We derive the Bethe ansatz equations describing the complete spectrum
of the transition matrix of the partially asymmetric exclusion process
with the most general open boundary conditions. For totally asymmetric
diffusion we calculate the spectral gap, which characterizes the
approach to stationarity at large times. We observe boundary induced crossovers 
in and between massive, diffusive and KPZ scaling regimes.  
\end{abstract}


\maketitle

The partially asymmetric simple exclusion process (PASEP) describes
the asymmetric diffusion of particles along a one-dimensional chain
with $L$ sites. It is one of the most studied models of
non-equilibrium statistical mechanics, see \cite{Derrida98,Schuetz00}
for recent reviews. This is in part due to the fact that is one of the
simplest lattice gas models, but also because of its applicability to
molecular diffusion in zeolites \cite{HahnKK96}, bioploymers
\cite{biopolymer}, traffic flow \cite{ChowdSS00} and other
one-dimensional complex systems \cite{Privman}. 

At large times the PASEP exhibits a relaxation towards a
nonequilibrium stationary state. An interesting feature of the PASEP
is the presence of boundary induced phase transitions \cite{Krug91}. 
In particular, in an open system with two boundaries at which
particles are injected and extracted with given rates, the bulk
behaviour in the stationary state is strongly dependent on the
injection and extraction rates. Over the last decade many stationary
state properties of the PASEP with open boundaries have been
determined exactly
\cite{Derrida98,Schuetz00,DEHP,gunter,EsslerR95,PASEPstat}.  
On the other hand, much less is known about its dynamics. This is in
contrast to the PASEP on a ring for which exact results using Bethe's
ansatz have been available for a long time \cite{dhar,BAring}. 
For open boundaries there have been several studies of dynamical
properties based mainly on numerical and phenomenological methods 
\cite{numerics,DudzS00}. In this Letter we employ Bethe's ansatz to
obtain exact results for the approach to stationarity at large times
in the PASEP with open boundaries. Upon varying the boundary rates, we
find crossovers in massive regions, with dynamic exponents $z=0$, and
between massive and scaling regions with diffusive ($z=2$) and KPZ
($z=3/2$) behaviour.

The dynamical rules of the PASEP are as follows. At any given time
$t$ each site is either occupied by a particle or empty and the system
evolves subject to the following rules. In the bulk of the system
($i=2,\ldots,L-1$) a particle attempts to 
hop one site to the right with rate $p$ and one site to the left with
rate $q$. The hop is executed unless the neighbouring site is
occupied, in which case nothing happens. On the first and last sites
these rules are modified. If site $i=1$ is empty, a particle may 
enter the system with rate $\alpha$. If on the other hand site $1$ is
occupied by a particle, the latter will leave the system with rate
$\gamma$. Similarly, at $i=L$ particles are injected and extracted
with rates $\delta$ and $\beta$ respectively. With every site $i$ we
associate a Boolean variable $\tau_i$, indicating whether a particle
is present ($\tau_i=1$) or not ($\tau_i=0$). The state of the system
at time $t$ is then characterized by the probability distribution
$P_t(\tau_1,\ldots,\tau_L)$. The time evolution of $P_t$ occurs
according to the aforementioned rules and as a result is subject to the
master equation    
\be
\frac{\d P_t}{\d t} = M P_t.
\label{eq:Markov}
\ee
Here $M$ is the PASEP transition matrix whose eigenvalues have
non-positive real parts. The large time behaviour of the PASEP is
dominated by the eigenstates of $M$ with the largest real parts of the 
corresponding eigenvalues.

{\it Bethe's Ansatz:}
It is well known that the transition matrix $M$ is related to the
Hamiltonian $H$ of the open spin-1/2 XXZ quantum spin chain through a
similarity transformation $M = - \sqrt{pq}\, U_{\lambda} H U_{\lambda}^{-1}$ where $H$ is
given by \cite{EsslerR95} 
\bea
H &=& -\frac12 \sum_{j=1}^{L-1} \left[
  \sigma_j^{\rm x}\sigma_{j+1}^{\rm x} + \sigma_j^{\rm
    y}\sigma_{j+1}^{\rm y} -\Delta \sigma_j^{\rm z}\sigma_{j+1}^{\rm
    z} \right.\nonumber\\
&& \hphantom{-\frac{\sqrt{pq}}{2} }
\left. {}+ h
    (\sigma_{j+1}^{\rm z}-\sigma_j^{\rm z}) +\Delta \right] +B_1 + B_L.
\eea
The parameters $\Delta$ and $h$, and the boundary terms $B_{1,L}$
are related to the PASEP transition rates by
\bea
\Delta&=& -\frac12(Q+Q^{-1}),\quad h=\frac12(Q-Q^{-1}),\quad
Q=\sqrt{\frac qp},\nonumber\\
B_L&=&\frac{\beta+\delta-(\beta-\delta)
\sigma_L^{\rm z} - \frac{2\beta}{\lambda Q^{L-1}}
\sigma_L^{+}- 2\delta\lambda Q^{L-1}\sigma_L^{-}}{2\sqrt{pq}} 
,\nonumber\\
B_1 &=&\frac{\alpha+\gamma+(\alpha-\gamma)
\sigma_1^{\rm z} - 2\alpha\lambda \sigma_1^{-} - \frac{2\gamma}{\lambda}
\sigma_1^{+}}{2\sqrt{pq}}.
\eea
Here $\lambda$ is a free parameter on which the spectrum does not
depend and $\sigma_j^{\pm} = (\sigma_j^{\rm x} \pm \i \sigma_j^{\rm
  y})/2$.  

Although it has been known for a long time that
$H$ is integrable \cite{GonzalesRuiz}, the off-diagonal terms in $B_1$
and $B_L$ have presented great difficulties in diagonalizing $H$ using
e.g. Bethe's ansatz. However, recently a breakthrough was achieved
\cite{Nepo02} in the case where the parameters satisfy a 
constraint, which in our notation reads 
\be
(Q^{L+2k} -1)(\alpha\beta - \gamma\delta Q^{L-2k-2}) =0.
\label{eq:constraint}
\ee
Here $k$ is an integer such that $|k|\leq L/2$. For a given $k$ this
constraint can be satisfied by choosing $Q$ to be an appropriate root
of unity, or by relating the boundary and bulk parameters such that the
second factor in (\ref{eq:constraint}) is zero. 
However, for generic values of the PASEP parameters
(\ref{eq:constraint}) can also be satisfied by choosing $k=-L/2$. For
this choice of $k$ we infer from \cite{NepoR03} that for even $L$ 
there is an isolated level with energy $E_0=0$, the ground
state energy of the PASEP. Furthermore, all excited levels are given
by
\bea
E=\alpha+\beta+\gamma+\delta+\sum_{j=1}^{L-1}\frac{\left(Q^2-1\right)^2
  z_j}{(Q-z_j)(Qz_j-1)},
\label{eq:pasep_en}
\eea
where the complex numbers $z_j$ satisfy the Bethe ansatz equations
\be
\left[\frac{z_jQ-1}{Q-z_j}\right]^{2L} K(z_j) =\prod_{l\neq j}^{L-1}
\frac{z_jQ^2-z_l}{z_j-z_lQ^2} \frac{z_jz_lQ^2-1} {z_jz_l-Q^2}.
\label{eq:pasep_eq}
\ee
Here $K(z) = \tilde{K}(z,\alpha,\gamma) \tilde{K}(z,\beta,\delta)$ and
\be
\tilde{K}(z,\alpha,\gamma) =
\frac{-\alpha z^2+Qz(Q^2-1+\alpha-\gamma)+\gamma Q^2}{\gamma Q^2
  z^2+Qz(Q^2-1+\alpha-\gamma)-\alpha}. 
\ee
In order to ease notations we have set, without loss of generality,
$p=1$ and hence $Q=\sqrt{q}$. We note that in the case of symmetric
diffusion $Q=1$ \r{eq:pasep_eq} reduce to the Bethe ansatz equations
derived in \cite{robin} by completely different means.
In order to determine the exact value of the spectral gap we have
analyzed (\ref{eq:pasep_en}) and (\ref{eq:pasep_eq}) in the limit
$L\rightarrow\infty$. To simplify the analysis, we will focus on the
case of total asymmetry $\gamma=\delta=0$, $Q\rightarrow 0$ in the
remainder of this Letter. 

{\sl Totally asymmetric exclusion (TASEP):}
After a rescaling $z \rightarrow Q z$ and
setting $\gamma=\delta=0$, the  $Q\rightarrow 0$ limit of equations
\r{eq:pasep_en} and \r{eq:pasep_eq} reads
\bea
&& \hspace{-5mm}E = \alpha+\beta + \sum_{l=1}^{L-1} \frac{z_l}{z_l-1},
\label{eq:tasep_en}\\
&&\hspace{-5mm}\left(\frac{(z_j-1)^2}{z_j}\right)^{L} = \left(z_j +
a\right) \left(z_j + b\right) \prod_{l\neq j}^{L-1} \left(
z_j-z_l^{-1}\right),
\label{eq:tasepBAE}
\eea
where $a=(1-\alpha)/\alpha$ and $b=(1-\beta)/\beta$. 
We define
$g_{\rm }(z) = \ln z/(z-1)^2$ and
$g_{\rm b}(z) = \ln z/(1-z^2) +
\ln\left(z+a\right) + \ln\left(z+b\right)$,
and consider the ``counting function'' \cite{deVegaW85}, 
\be
\i Y_L(z) = g_{\rm}(z) + \frac{1}{L} g_{\rm b}(z)
+ \frac{1}{L} \sum_{l=1}^{L-1} K(z_l,z),
\label{eq:logtasepBAE}
\ee
where $K(w,z) = -\ln w + \ln(1-w z).$
Equations \r{eq:tasepBAE} can now be written as
\be
Y_L(z_j) = \frac{2\pi}{L} I_j\quad (j=1,\ldots,L-1),
\label{eq:Z=I}
\ee
where the $I_j$ are integers. Each set of integers $\{I_j\}$ specifies
a particular excited state, and we find numerically that the first
excited state is obtained for the choice
\be
I_j = -L/2+j\quad {\rm for}\quad j=1,\ldots,L-1.
\label{eq:Idef}
\ee
The eigenvalue \r{eq:tasep_en} can be written as  
\be
E = \alpha+\beta + L \lim_{z\rightarrow 1} \left( \i\, Y_L'(z) -
  g'(z) - \frac{1}{L} g'_{\rm b}(z)\right).
\label{eq:EinY}
\ee

\begin{figure}[ht]
\begin{center}
\begin{picture}(145,115)
\put(0,0){\epsfig{file=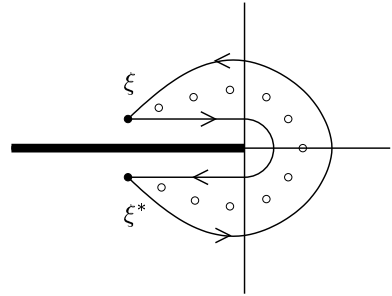}}
\end{picture}
\end{center}
\caption{Sketch of the contour of integration $C$ in \r{eq:sum2int}. The
  open dots correspond to the roots $z_j$ and $\xi$ is chosen close to
  $z_{L-1}$ and avoiding poles of $\cot(LY_L(z)/2)$.} 
\label{fig:contour}
\end{figure}

In order to derive an integral equation for $Y_L(z)$ in the limit
$L\rightarrow\infty$ we write, 
\be
\frac1L \sum_{j=1}^{L-1} f(z_j) = \oint_{C_1+C_2} 
\frac{dz}{4\pi\i}\ f(z)
Y_L'(z) \cot\left(\frac12 L Y_L(z)\right),
\label{eq:sum2int}
\ee
where $C=C_1+C_2$ is a contour enclosing all the roots $z_j$, $C_1$
being the ``interior'' and $C_2$ the ``exterior'' part, see
Fig.~\ref{fig:contour}. The contours $C_1$ and $C_2$ intersect in
appropriately chosen points $\xi$ and $\xi^*$. 
Using the fact that integration from $\xi^*$ to $\xi$ over the contour
formed by the roots is equal to half that over $C_2 - C_1$ we find
using \r{eq:sum2int},
\bea
\i\,Y_L(z) &=& g(z) + \frac{1}{L} g_{\rm b}(z) +\frac{1}{2\pi}
\int_{\xi^*}^{\xi} K(w,z) Y'_L(w) \d w \nonumber \\ 
&&+ \frac{1}{2\pi} \int_{C_1} \frac{K(w,z)Y'_L(w)}{1-\e^{-\i L
Y_L(w)}}\, \d w\nonumber\\
&&+ \frac{1}{2\pi} \int_{C_2} \frac{K(w,z)Y'_L(w)}{
\e^{\i L Y_L(w)}-1}\,\d w,
\label{eq:intY}
\eea
where we have chosen the branch cut of $K(w,z)$ to lie along the
negative real axis. The eigenvalue \r{eq:EinY} is obtained by
systematically expanding \r{eq:intY} in the system size using standard
methods.
We note that care has to be taken when there is a
stationary point of $Y_L(z)$ close to the contour of integration, in
which case an analysis similar to that in \cite{BAring} has to be
carried out.  

Let us briefly recall the stationary state phase diagram derived in
\cite{gunter,DEHP}. There are altogether four phases in the stationary
state at $t=\infty$:
(1) the low density phase for $\alpha<\beta<1/2$;
(2) the high density phase for $\beta<\alpha<1/2$;
(3) the coexistence line at $\beta=\alpha<1/2$;
(4) the maximal current phase at $\alpha,\beta>1/2$.
We now determine the scaling of the spectral gap in these regimes from
the Bethe ansatz equations. 

{\it Low and High Density Phases:}
Let us fix the end points $\xi^*$ and $\xi$ by 
\be
Y_L(\xi^*) = -\pi +\frac{\pi}{L},\qquad Y_L(\xi) = \pi -\frac{\pi}{L}.
\label{eq:xidef}
\ee
The integral over $C_1$ in \r{eq:intY} can be calculated by splitting
the contour into its upper and lower parts and expanding the integrand around
$\xi$ and $\xi^*$ respectively. 
Expanding in inverse powers of $L$, i.e.,
\be
Y_L(z) =\sum_{n=0} L^{-n}Y_n(z),\qquad \xi = z_{\rm c} + \sum_{n=1}\delta_n
L^{-n},
\label{eq:expand}
\ee
and assuming that $-a<z_{\rm c}$ and $-b<z_{\rm c}$, we find from
equation \r{eq:intY} to ${\cal O}(L^{-1})$ that,
\bea
Y_{0}(z) &=& -\i \ln \left[ - \frac{z}{z_{\rm c}}
\left(\frac{1-z_{\rm c}}{1-z}\right)^2 \right],
\label{eq:Z0sol}
\\
Y_1(z) &=& - \i \ln\left[ -\frac{z}{z_{\rm c}} \frac{1-z_{\rm c}^{2}}{1-z^{2}}
\left[\frac{z_{\rm c}-z_{\rm c}^{-1}}{z-z_{\rm c}^{-1}}\right]^{\nu_1}
\frac{z+a}{z_{\rm c}+a} \frac{z+b}{z_{\rm 
    c}+b}\right]\nonumber\\
&& -\i\ln \left(ab (-z_{\rm c})^{\nu_1}\right)
\label{eq:Z1sol},
\eea
where $\nu_1 = -Y_0'(z_{\rm c})\delta_1/\pi$.
The values of $\nu_1$ and $z_{\rm c}$ follow from (\ref{eq:xidef}) to be
$\nu_1 =2$, $z_{\rm c}=-1/\sqrt{ab}$. Substituting 
these values into \r{eq:EinY} we obtain the gap \r{eq:E_MI}, which is
of order ${\cal O}(1)$ in the limit $L\to\infty$.

If $-b > -1/\sqrt{ab}$ the point $-b$ lies inside the contour formed
by the roots, see Fig~\ref{fig:contour}, giving rise to 
a different solution for $Y_1(z)$.
Comparing again with condition (\ref{eq:xidef}) we find in this case
that $\nu_1=3$ and $z_{\rm c} = -a^{-1/3} = -1/\sqrt{ab_{\rm c}}$, 
resulting in the spectral gap given in \r{eq:E_MII},
which is independent of $\beta$ and again of order ${\cal O}(1)$ in
the limit $L\to\infty$. A similar analysis is made when $-a >
-1/\sqrt{ab}$. As the spectral gap is ${\cal O}(1)$ in the
low and high density phases, the correlation length is finite and
these phases are therefore massive.

{\it Coexistence Line:}
Subleading corrections can be obtained by taking higher order terms
into account in 
\r{eq:expand}. After a lengthy
calculation we find that the gap vanishes like $L^{-2}$
on the coexistence line $\beta=\alpha$, with a constant of
proportionality given by \r{eq:E_EW}.

{\it Maximal Current phase:}
When $\beta>\beta_{\rm c}$ and $\alpha\rightarrow 1/2$, the value of
$z_{\rm c}$ where the contour closes approaches $z_{\rm c}=-1$ and a 
straightforward expansion of the last two terms in \r{eq:intY} breaks
down as $Y'_L(\xi^*) \approx Y'_0 (-1)= 0$. 
A further complication is  the singularity in $K(w,z)$ at $w=z=z_{\rm c}=-1$.
To proceed one expands around
$z_{\rm c}$ defined by $Y'_L(z_{\rm c})=0$
\cite{BAring,PovoPH03}. This gives rise to an expansion of
$Y_L(z)$ in powers of $L^{-1/2}$ and one finds in lowest order that
the energy gap vanishes as $L^{-3/2}$. The prefactor can 
only be determined numerically.

We now summarize our results. We have used Bethe's ansatz to
diagonalize the PASEP transition matrix $M$ for
arbitrary values of the rates $p$, $q$, $\alpha$, 
$\beta$, $\gamma$ and $\delta$ that characterize the most general
PASEP with open boundaries. 
The resulting Bethe ansatz equations \r{eq:pasep_en}, \r{eq:pasep_eq}
describe the \emph{complete} excitation spectrum of $M$ and are one of our
main results. We have carried out a detailed analysis of the Bethe
ansatz equations for the simplified case of the TASEP and
determined the exact asymptotic behaviour of the spectral gap for
large lattice lengths $L$. This gap determines the long time ($t\gg
L$) dynamical behaviour of the TASEP. We emphasize that care has to be 
taken regarding time scales, and that our results below are not valid
at intermediate times $t\approx L$ where the system behaves as
for periodic boundary conditions \cite{Schuetz00}.

\begin{figure}[h!]
\begin{center}
\begin{picture}(160,120)
\put(0,-20){\epsfig{file=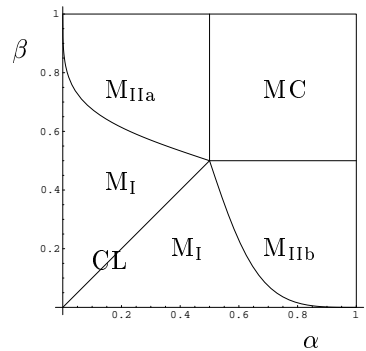}}
\end{picture}
\end{center}
\caption{Dynamic phase diagram of the TASEP. M$_{\rm I}$, M$_{\rm
    IIa}$ and M$_{\rm IIb}$ are massive phases, CL 
denotes the critical coexistence line and MC the critical maximal
current phase.}
\label{fig:phase}
\end{figure}

We found that there are three regions in parameter space where the
spectral gap is finite and the stationary state is approached
exponentially fast, and one region and a line where the gap vanishes
as $L\rightarrow\infty$. The resulting dynamical phase diagram is
shown in Figure~\ref{fig:phase}.  In order to parametrize the phases
we define $\beta_{\rm c} = (1+a^{-1/3})^{-1}$ and $\alpha_{\rm c} =
(1+b^{-1/3})^{-1}$. 
The values of the spectral gap in the various regions of the phase diagram
of the TASEP are as follows:

\noindent{{\it Massive Phase M$_{\rm I}$:
$\alpha< \alpha_{\rm c}$, $\beta<\beta_c$, $\alpha\neq\beta$}}
\be
-E_1 = \alpha+\beta - \frac{2}{1+\sqrt{ab}} +
\mathcal{O}(L^{-2}),
\label{eq:E_MI}
\ee
The spectral gap does not vanish as $L\rightarrow\infty$ and hence
implies a finite correlation length and exponential approach to stationarity.\\
\noindent{{\it Low Density Phase M$_{\rm IIa}$: 
$\alpha< 1/2$, $\beta>\beta_{\rm c}$}}
\be
-E_1 = \alpha+\beta_{\rm c} - \frac{2}{1+\sqrt{ab_{\rm c}}} +
\mathcal{O}(L^{-2}),
\label{eq:E_MII}
\ee
Note that in this phase the spectral gap is independent of
$\beta$. The behaviour in the high-density phase M$_{\rm IIb}$:
$\beta<1/2$, $\alpha_{\rm  c}<\alpha$ is obtained by the exchange
$\alpha\leftrightarrow\beta$. 

\noindent{\it Coexistence Line (CL): $\beta=\alpha<1/2$.} 
\be
-E_1 = \frac{\pi^2 \alpha(1-\alpha)}{1-2\alpha} L^{-2} +
\mathcal{O}(L^{-3}).
\label{eq:E_EW}
\ee
We thus find a dynamic exponent $z=2$ corresponding to 
diffusive behaviour.

\noindent{\it Maximal Current Phase (MC): $\alpha,\beta>1/2$.}
\be
-E_1 \approx 3.56 L^{-3/2} + \mathcal{O}(L^{-2}).
\ee
In this phase, which coincides with the stationary maximum current
phase, we find a KPZ \cite{KPZ}  dynamic exponent $z=3/2$. The gap is
smaller than that of the periodic case where it is found that $-E_1 =
6.509\ldots L^{-3/2}$ \cite{BAring,GoliM04}. 
We note that the subdivision of the massive high and low density
phases is different from the one suggested on the basis of stationary
state properties in \cite{gunter}.

{\it Discussion:}
It is known \cite{NeergN95} that by varying the bulk hopping
rates one can induce a crossover between a diffusive Edwards-Wilkinson
(EW) scaling regime \cite{EW} with dynamic exponent $z=2$ and a KPZ regime
\cite{KPZ} with $z=3/2$. In this letter we have shown using exact
methods that a crossover between phases with $z=2$ and $z=3/2$ occurs
in the case where the bulk transition rates are kept constant, but the
boundary injection/extraction rates are varied. As shown in
\cite{DudzS00} the diffusive relaxation ($z=2$) is of a different
nature than in the EW regime and is in fact due to the unbiased random
walk behaviour of a shock (domain wall between a low and high density
region). Our results \r{eq:E_MI} and \r{eq:E_EW} for
the massive phase $M_{\rm I}$  and the coexistence line agree with the
relaxation time calculated in the framework of a
domain wall theory (DWT) model in \cite{DudzS00}. This is in contrast to
the massive phases $M_{\rm II}$, where the exact result \r{eq:E_MII}
differs from the DWT prediction. An interesting open question is
whether it is possible to understand \r{eq:E_MII} in a generalized DWT
framework. 

The Bethe ansatz equations \r{eq:pasep_en}, \r{eq:pasep_eq}
allow for the exact determination of further spectral gaps. We find that the
eigenvalue of the transition matrix with the next largest real part
is complex, which leads to interesting oscillatory behaviour at large
times. The dynamical phase diagram for the general PASEP is expected
to be significantly richer, and its analysis is under way.

The condition \r{eq:constraint} is a reflection of the non-semi\-simplic\-i\-ty
of an underlying Temperley-Lieb algebra with two additional boundary
generators \cite{magic}. Remarkably the PASEP satisfies this
constraint for arbitrary values of its parameters. Generically,
non-semisimplicity implies certain symmetries in the spectrum, and the
physical consequences of these are currently under investigation. 

\acknowledgments
We are grateful to G.M. Sch\"utz for very helpful discussions. This 
work was supported by the ARC (JdG) and the EPSRC under grant
GR/R83712/01 (FE).

\end{document}